\documentclass[conference]{IEEEtran}
\IEEEoverridecommandlockouts
\usepackage{cite}
\usepackage{amsmath,amssymb,amsfonts}
\usepackage{algorithmic}
\usepackage{graphicx}
\usepackage{textcomp}
\usepackage{xcolor}
\usepackage{url}
\usepackage{hyperref}
\usepackage{multirow}
\usepackage{arydshln}
\def\BibTeX{{\rm B\kern-.05em{\sc i\kern-.025em b}\kern-.08em
    T\kern-.1667em\lower.7ex\hbox{E}\kern-.125emX}}
\begin{document}

\title{Restorative Speech Enhancement: A Progressive Approach Using SE and Codec Modules} %for Noise and Reverberation Reduction}

\author{
Hsin-Tien Chiang$^{1*}$, Hao Zhang$^{2}$, Yong Xu$^{2}$, Meng Yu$^{2}$, Dong Yu$^{2}$\\
\IEEEauthorblockA{$^{1}$ The University of Texas at Dallas, Richardson, TX, USA\\
$^{2}$ Tencent AI Lab, Bellevue, WA, USA}\\
\thanks{$^{*}$This work was done during an internship at Tencent AI Lab.}
}

\maketitle

\begin{abstract}
In challenging environments with significant noise and reverberation, traditional speech enhancement (SE) methods often lead to over-suppressed speech, creating artifacts during listening and harming downstream tasks performance. To overcome these limitations, we propose a novel approach called Restorative SE (RestSE), which combines a lightweight SE module with a generative codec module to progressively enhance and restore speech quality. The SE module initially reduces noise, while the codec module subsequently performs dereverberation and restores speech using generative capabilities. We systematically explore various quantization techniques within the codec module to optimize performance. Additionally, we introduce a weighted loss function and feature fusion that merges the SE output with the original mixture, particularly at segments where the SE output is heavily distorted. Experimental results demonstrate the effectiveness of our proposed method in enhancing speech quality under adverse conditions. Audio demos are available at: \url{https://sophie091524.github.io/RestorativeSE/}.
\end{abstract}


\section{Introduction}
Speech enhancement (SE) is important for improving listening experiences and enhancing the performance of various speech processing applications, such as automatic speech recognition systems, telecommunications, and hearing assistance devices. In real-world scenarios, environmental noise and room reverberation can severely degrade the quality of speech, making robust SE methods essential.
\par SE has been formulated as a supervised learning problem, with methods broadly categorized into time-frequency (T-F) domain and time-domain approaches. In T-F domain methods, speech signals are transformed into a T-F representation using short-time Fourier transform (STFT). Enhancement targets in this domain include masking-based approaches, such as the ideal binary mask (IBM) \cite{wang2005ideal}, ideal ratio mask (IRM) \cite{narayanan2013ideal}, and complex ratio mask (CRM) \cite{williamson2015complex}, as well as mapping-based approaches that directly estimate the spectral representation of clean speech \cite{lu2013speech,xu2013experimental}. Alternatively, time-domain methods estimate the clean speech signal directly from the raw waveform \cite{fu2017raw, pandey2019tcnn,defossez2020real}. However, most of these methods may lead to unwanted speech distortions \cite{wang2019bridging}.
\par Recently, there has been a growing interest in leveraging generative models for SE. These models, including generative adversarial networks (GANs) \cite{pascual2017segan,su2021hifi}, variational autoencoders (VAEs) \cite{fang2021variational, bie2022unsupervised}, flow-based models \cite{nugraha2020flow}, and diffusion probabilistic models \cite{lu2022conditional,lemercier2023storm}, have demonstrated superior noise handling and generalization capabilities \cite{wang2024selm}. However, under adverse conditions, such generative models might be prone to confusing phonemes, leading to the generation of unintended vocalizing artifacts \cite{richter2023speech,lemercier2023storm}.
%Additionally, research has also explored the use of vocoders
%\cite{wang2024selm} regenerates the discretized WavLM features and takes HiFi-Gan to recover enhanced speech.
%\cite{irvin2023self} directly predicts the clean speech through the vocoder from a noisy representation extracted from a large self-supervised learning (SSL) models. 
\par In this paper, we propose a novel framework called \textquotedblleft Restorative SE\textquotedblright (RestSE), which combines a lightweight SE module with a generative codec module to progressively enhance and restore speech signal. The contribution of this paper is as follows: (1) we design a progressive learning pipeline that separates the SE process into two stages: denoising followed by dereverberation. The denoising stage employs a lightweight SE module to suppress noise, while the dereverberation stage employs a generative codec module to remove reverberation and restore the speech signal; (2) we conduct a comprehensive investigation of quantization techniques within the codec module to optimize performance. Specifically, we explore configurations of scalar quantization (SQ) and vector quantization (VQ) and propose a new design that sequentially applies SQ followed by residual VQ, resulting in improved restoration quality; and (3) we introduce a weighted loss function and feature fusion that merges the SE output with the original mixture to ensure robust restoration.

\section{Proposed Approach}
\label{sec:method}
Let $s(t)$, $x(t)$, $n(t)$ and $h(t)$ represent the dry clean speech, reverberant speech, background noise and room impulse response (RIR) function, respectively. The noisy-reverberant mixture $y(t)$ can be expressed as 
\begin{equation}
y(t) = x(t) + n(t) = s(t)*h(t)+n(t)
\end{equation}
where $*$ denotes the convolution operator.

\begin{figure}[t!]
%\begin{minipage}[b]{1\linewidth}
  \centering
  \includegraphics[width=1\linewidth]{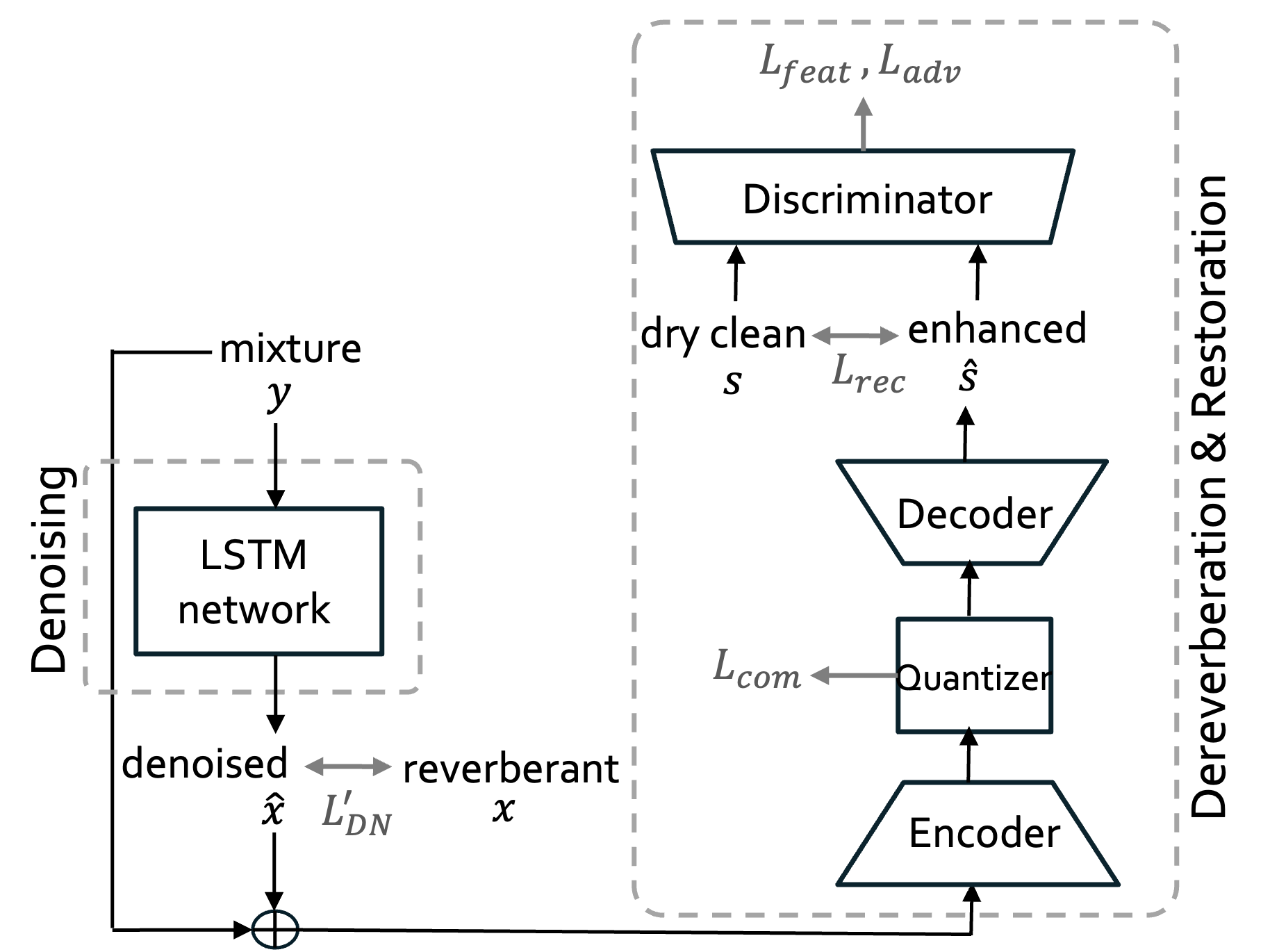} % Adjust width if needed
%\end{minipage}
\caption{The proposed RestSE framework, illustrating the progressive pipeline with two sequential stages: DN and DR\&RST.}
\label{fig:pipline}
\end{figure}

The proposed RestSE approach is structured into a progressive pipeline with two sequential stages, as illustrated in Fig. \ref{fig:pipline}. The first stage, referred to as the denoising (DN) stage, employs a lightweight SE module to reduce background noise only, with the target being the reverberant speech $x$. The second stage, termed the dereverberation and restoration (DR\&RST) stage, utilizes a codec module to remove reverberation and restore speech signal, ultimately aiming to recover the dry clean speech $s$. The motivation for using a progressive SE pipeline arises from the difficulty of a standalone SE module to handle both DN and DR effectively, as these tasks have different characteristics \cite{zhao2017two,zhao2018two}. Wang's recent study \cite{wang2023unifying} showed that performing SE in the codec space reduces uncertainty and ambiguity in restoration. Based on this, we integrate a codec module in the DR\&RST stage to improve speech restoration performance.

\subsection{DN stage}
To suppress noise, we employ the T-F masking framework with CRM estimated using long-short term memory (LSTM) architecture. The LSTM network consists of three hidden layers, each containing 300 units. The loss function is calculated using the combination of SI-SDR \cite{le2019sdr} loss in the time domain and the $L_1$ loss on the spectrum magnitude difference between the denoised signal and the target reverberant signal: 
%We use 512-point FFT, with 32ms hann window and 16ms hop siz 
\begin{equation}
L_{DN} = -\text{SI-SDR}(\hat{x}, x) + \lambda \cdot L_1(\vert \hat{X} \vert, \vert X \vert)
\label{eqn:lossDN}
\end{equation}
where $\lambda$ is set to 1000 to balance the value range of the two losses, with $\vert \hat{X} \vert$ and $\vert X \vert$ representing the magnitudes of the denoised signal and the target reverberant signal, respectively. The LSTM network is pretrained using this loss function and then jointly trained with the codec module.

\subsection{DR \& RST stage}
%After suppressing background noise, the original task is reduced to recovering the dry clean speech $s$ from the reverberant speech $x$. 
%We utilize a codec module for dereverberation and speech restoration, as \cite{wang2023unifying} has demonstrated the effectiveness of speech codecs in SE under both simulated and real-world conditions.
We employ a codec module for dereverberation and speech restoration, following \cite{wang2023unifying}, which demonstrated the effectiveness of speech codecs in SE for both simulated and real-world scenarios.
These codec modules generally consist of an encoder, a quantizer, and a decoder.

\subsubsection{Encoder and decoder}
We adopt the same SEANet architectures as SoundStream \cite{zeghidour2021soundstream} and Encodec \cite{defossez2022high} for encoder and decoder. The encoder and decoder are then initialized with a pretrained SoundStream model. Note that the encoder and decoder are kept fixed, and our primary focus is on exploring different quantization techniques.

\subsubsection{Quantization} 
Quantization are generally categorized into two groups: SQ \cite{balle2016end,yang2024simplespeech} and VQ \cite{zeghidour2021soundstream, defossez2022high,yang2023hifi}. In the SQ method, each sample is quantized independently, without considering correlations between different dimensions. This approach is advantageous due to its simplicity and lower computational complexity. Specifically, the SQ module quantizes the output of the encoder, denoted as $z \in R^{F \times D}$, where $F$ represents the number of frames and $D$, set to 256, denotes the dimensionality of each vector. The quantization process maps the output into a fixed scalar space:
\begin{equation}
z = \tanh(z)
\end{equation}
\begin{equation}
\quad r = \text{round}(z \times K)/K
\end{equation}
where $K$ is a hyper-parameter that determines the scope of the scalar space, and $M$ is set to 8. The \emph{tanh} function is initially applied to compress the feature values into the range $[-1, 1]$, and the \emph{round} operation reduces the value of range into 2$\times M$+1 different values. %To allow gradient backpropagation through the rounding step, we employ a straight-through estimator similar to that used in VQ-VAE \cite{van2017neural}. The \emph{tanh} function is initially applied to compress the feature values into the range $[-1, 1]$, and the \emph{round} operation reduces the value of range into 2$\times M$+1 different values.

\par In contrast, VQ operates by matching an input vector to the closest entry in a codebook and is effective at compressing highly correlated data and achieving better reconstruction quality in audio codecs. However, VQ comes with increased complexity, particularly as the vector dimensions and the size of the codebook grow. Specifically, VQ learns a codebook of $N$ vectors to encode each $D$-dimensional frame of the encoder output $z$, which is then mapped to a sequence of one-hot vectors with shape $F \times N$, where $N$ equals to 1024.
 
%VQ operates by matching an input vector to the closest entry in a codebook, processing multiple dimensions simultaneously. %The codebook entry selected for each input is updated using an exponential moving average with a decay of 0.99, and entries that are not used are replaced with a candidate sampled from the current batch For all of our models, we use at most 32 codebooks (16 for the 48 kHz models) with 1024 entries each, e.g

\par In this study, we conduct a comprehensive exploration of various quantization techniques. Beyond SQ, we investigate the commonly used Residual VQ (RVQ), which recursively quantizes residuals using distinct codebooks after an initial quantization step. We also examine a residual version of SQ, termed Residual SQ (RSQ). Additionally, we propose hybrid strategies that integrate SQ and RVQ in different configurations, including the sequential application of SQ followed by RVQ (SQ-RVQ), a grouped configuration of SQ-RVQ, and a parallel configuration where SQ and RVQ are combined through a weighted sum (SQ $\parallel$ RVQ). For a fair comparison, the number of quantization modules $N_q$ is fixed at 8 across all residual-based methods (i.e., RSQ, RVQ, and SQ-RVQ). 

\subsubsection{Training objective}
The training loss consists of four parts: (1) reconstruction loss, $L_{rec}$, which utilizes multi-resolution STFT loss across both full-band and sub-band scales \cite{yang2021multi}; (2) adversarial loss, $L_{adv}$, computed using hinge loss based on multi-scale STFT discriminators \cite{defossez2022high}; (3) $L_{feat}$, applying the $L_1$ feature loss \cite{kumar2019melgan}; and (4) commitment loss, $L_{com}$, applied between the output of the encoder, and its quantized representation \cite{zeghidour2021soundstream, defossez2022high}. The respective weightings for these losses are set to 1 for both the reconstruction and adversarial losses, 20 for the feature loss, and 10 for the commitment loss.

\subsection{Enhancing RST with weighted loss and feature fusion}
However, we found that the DN stage not only reduces noise but also suppresses crucial speech content within the mixture, posing a challenge for the codec model in the DR stage to fully recover the enhanced speech due to the loss of essential information before it reaches the codec input. To address this issue, we introduce a weighted loss function for the LSTM network and apply feature fusion that merges the denoised speech with the original mixture. The result of the fusion is then used as input for the codec to compensate for the missing information in the enhanced speech.

\subsubsection{Weighted loss}
We propose a weighted loss function that emphasizes regions where the speech signal might be overly suppressed. 
The weighting factor $\alpha$ is obtained as:
\begin{equation}
\alpha = 1 + \frac{|\Delta X'|}{\max(|\Delta X'|)} \cdot M
\end{equation}
where $M$ is a binary mask that applies the weight selectively to regions where the reference magnitude \(X\) exceeds a small predefined threshold (i.e., $1 \times 10^{-8}$). This threshold ensures that the weighting factor \(\alpha\) only influences significant spectral components, preventing unnecessary emphasis on low-energy regions.
$\Delta X'$ is defined as:
\begin{equation}
\Delta X' =
\begin{cases}
2 \cdot \Delta X & \text{if } \Delta X < 0 \\
\Delta X & \text{otherwise } %\Delta X \geq 0
\end{cases}
\end{equation}
with $\Delta X=\vert \hat{X} \vert - \vert X \vert$ denoting the difference between the denoised spectrum magnitude $\vert \hat{X} \vert$ and the reverberant spectrum magnitude $\vert X \vert$.
\par 
Finally, this weighting factor $\alpha$ is integrated into the \(L_1\) loss term in (\ref{eqn:lossDN}), resulting in the following modified loss function:
\begin{equation}
L'_{DN} = -\text{SI-SDR}(\hat{x}, x) + \lambda \cdot \alpha \cdot L_1(\vert \hat{X} \vert, \vert X \vert)
\end{equation}
Through this, the weighted loss doubles the weight of the negative differences, thereby prioritizing regions where the speech components might have been overly suppressed.

\subsubsection{Feature fusion layer integration}
Previous works \cite{fan2020gated, zhu2023joint} have tackled speech distortion by combining noisy spectral features with enhanced ones, leveraging clean speech data from the enhanced features while capturing complementary details from the noisy features. Building on this strategy, we propose a feature fusion strategy that utilizes a linear layer to process both denoised speech and the mixture's magnitudes to create a unified representation. Which is then fed to the codec module for further enhancement. Incorporating the mixture allows for the preservation of fine structures from the noisy features, thereby reducing over-suppression and improving speech restoration. 
 
%The fused speech is a a linear layer which takes the the denoised speech and mixture as input, take both the magnitude and passed through a linear layer to obtained the fused representation, which later on takes the phase of denoised speech to obtain the fused waveform. so that not only remove the noise signals from the enhanced features, but also learn the raw fine structures from the noisy features so that it can alleviate the speech distortion.
%The magnitude are used as features used to enhance the output for the denoising stage of the over suppressed problem. Let $y_{DN}$ and $x$ denote the mixture and the denoised signals in the time domain, while $Y_{DN} =[Y_{DN}(1), ...Y_{DN}(F)]$ and $X=[X(1), ...X(F)]$ represents the magnitude for $y_{DN}$ and $x$, respectively in which F is the number of frequency bins. The magnitude $Y_{DN}$ of the denoising stage $y_{DN}$  is concatenated with the magnitude $X$ of the mixture, $x$ and then feed into the post-processor to obtain the residual feature

\section{Experimental setup}
\label{sec:experiments}
\subsection{Datasets and settings}
The experiments are conducted using a combination of AISHELL-2 \cite{du2018aishell} and LibriSpeech \cite{panayotov2015librispeech} datasets. For room impulse responses (RIRs), we generate 10,000 RIRs using the image method \cite{allen1979image} with random room characteristics and reverberation times (RT60) randomly selected within the range of 0 to 0.6 seconds. 
During training, for each training sample, we randomly select a speech sample and an RIR for generating the target speech and reverberated signal. Environmental noises are then added to the speech with signal-to-noise-ratio (SNR) uniformly distributed between -6 and 6 dB. In total, we generated 90000 and 6000 utterances for training and validation, respectively. Each speech segment length set to 6 seconds.
For testing, we generated a test set with three distinct SNR levels: -5, 0, and 5 dB, each containing 100 utterances. These test utterances and RIRs were not included in the training process, ensuring a rigorous evaluation of the model's performance on unseen data.

\par We trained the model using 4 V100 GPUs, with a total batch size of 16 over 50 epochs. The models were optimized using the Adam optimizer, with a learning rate of $2 \times 10^{-4}$.

\subsection{Evaluation metrics}
To assess the performance of our models, we utilize several key evaluation metrics, including perceptual evaluation of speech quality (PESQ) and short-time objective intelligibility (STOI). In addition, we incorporate the overall quality (OVRL) metric from DNSMOS P.835 \cite{reddy2022dnsmos}. 
Notably, DNSMOS is useful for generative models, as it effectively captures their ability to address issues such as noise leakage and excessive suppression \cite{wang2024selm}. Therefore, we emphasize the OVRL results to highlight the effectiveness of our approach.
%We evaluate our models using key metrics, including perceptual evaluation of speech quality (PESQ), short-time objective intelligibility (STOI), and the overall quality (OVRL) metric from DNSMOS P.835 \cite{reddy2022dnsmos}.
%Notably, DNSMOS is useful for generative models, as it captures issues like noise leakage and excessive suppression\cite{wang2024selm}. Therefore, we emphasize the OVRL results to highlight the effectiveness of our approach.

\section{Results}
\label{sec:results}
\subsection{Effectiveness of progressive learning pipeline}
We compare our approach against the codec-based method by Wang et al. \cite{wang2023unifying}, which uses supervised enhancement in the code token space to predict clean tokens, followed by decoding with a pretrained speech decoder to restore clean speech. We also evaluate the effectiveness of progressive learning through three configurations: (1) a codec module for joint DN and DR, denoted as codec (DN+DR); (2) a progressive learning pipeline with a randomly initialized LSTM network, referred to as LSTM (rand ini), that is then jointly trained with the codec module; and (3) a progressive pipeline using a pretrained LSTM, referred to as LSTM (pretrained), followed by joint training with the codec module (our proposed method). In all codec models, quantization is performed using SQ-RVQ, which yielded the best results compared to other techniques, as shown in Table \ref{table:diffQ}. It is also worth noting that we evaluated the proposed pipeline with both SE and codec targeting dry clean speech. However, the results were comparable to the codec (DN+DR) configuration, suggesting that the lightweight SE module may not be effective for joint DN and DR tasks. %(\textbf{Mention that we also utilized the proposed pipeline but set the target of both SE and codec as dry clean, results are similar to that of codec (DN+DR) since the light weighted SE if not proper for joint DN and DR task})

\begin{table}[t]
    \centering
    \caption{Comparison of progressive learning pipelines and baseline methods. Values in parentheses represent intrusive metrics with reverberant speech as the reference.}
    \label{table:baseline}
    %\scalebox{0.89}{ %
    \begin{tabular}{lcccc}
    \hline
    & PESQ $\uparrow$ & OVRL $\uparrow$ & STOI $\uparrow$ & Params (M) \\
    \hline
    Mixture & 1.143 & 1.207 & 0.513 & - \\
            & (1.383) & & (0.702) & \\
    \hline
    Wang et al. \cite{wang2023unifying} & 1.286 & 3.01 & 0.69 & 54.43 \\
    \hline
    codec (DN+DR) & 1.35 & 3.02 & 0.7 & 15.04 \\
    \hline
    LSTM (rand init) & 1.159 & 1.323 & 0.52 & \multirow{3}{*}{2.92 + 15.04} \\
         $\Downarrow$            & (1.426) & & (0.816) & \\
    \cdashline{1-4}
    codec & 1.371 & 2.998 & 0.71 & \\
    \hline
    LSTM (pretrained) & 1.281 & 1.892 & 0.562 & \multirow{3}{*}{2.92 + 15.04} \\
          $\Downarrow$             & (2.133) & & (0.816) & \\
    \cdashline{1-4}
    codec (RestSE) & \textbf{1.443} & \textbf{3.046} & \textbf{0.735} & \\
    \hline
    \end{tabular}  
    %}
\end{table}

\par We draw several observations from the Table. First, the progressive learning pipeline with LSTM (rand init) shows comparable performance to codec (DN+DR). We attribute this to the minimal improvement provided by LSTM (rand init) in DN, placing a greater burden on the codec module to manage both DN and DR. In contrast, using LSTM (pretrained) demonstrates significant improvement over LSTM (rand init), validating the effectiveness of the proposed progressive learning approach. Second, compared to Wang et al., our approach not only outperforms across all objective metrics but also results in a model size that is nearly one-third of theirs, demonstrating the efficiency and effectiveness of our progressive learning pipeline.

\subsection{Comparison of different quantization approches}
We observed that incorporating SQ, whether alone or combined with RVQ, consistently outperforms relying solely on RVQ. This is due to SQ's ability to independently quantize each sample, minimizing quantization error at a granular level and providing a precise initial approximation of the signal. Additionally, RVQ excels in refining residuals, which typically contain finer details, due to its effectiveness in handling complex structures. This is evidenced by the underperformance of RSQ compared to RVQ in residual quantization. Hence, the proposed SQ-RVQ configuration leverages the strengths of both techniques, with SQ capturing the primary components and RVQ refining the finer details, resulting in superior performance over either method alone. We also explored a group-based SQ-RVQ mechanism but observed a decline in performance. Additionally, RFSQ and RLFQ produced suboptimal results, likely because they were originally designed for images rather than speech, indicating that further adaptation is needed for better outcomes.

\subsection{Evaluation of weighted loss and feature fusion}
Table \ref{table:oversuppress} shows the impact of the weighted loss function and feature fusion on our codec model, which uses SQ-RVQ quantization for its superior performance. Removing the weighted loss function during the DN stage results in declines across all metrics. While adding feature fusion as input for the DR stage causes slight decreases in PESQ and STOI, it improves the OVRL score, highlighting its advantage in enhancing perceived speech quality despite trade-offs with objective measures.

%The results in Table \ref{table:oversuppress} demonstrate the impact of the weighted loss function and feature fusion layer on the performance of our codec model, which utilizes the SQ-RVQ quantization due to its superior performance. Removing the weighted loss function during the DN stage leads to declines across all objective metrics. Furthermore, although adding the feature fusion layer in the DR stage leads to minor decreases in PESQ and STOI, it improves the OVRL score, underscoring the advantage of the fusion layer in enhancing human perception quality, even if it results in trade-offs with objective metrics. %reveal the gap betwwen objective measures and human perception quality. ndicating a trade-off 

\begin{table}[t]
    \centering
    \caption{Comparison of Quantization Techniques}
    \label{table:diffQ}
    \begin{tabular}{lccc}
    \hline
    & PESQ $\uparrow$& OVRL$\uparrow$ & STOI$\uparrow$ \\
    \hline
    SQ \cite{balle2016end,yang2024simplespeech} & 1.422 & 3.008 & 0.727\\
    RSQ & 1.424 & 3.035 & 0.731\\
    RVQ \cite{zeghidour2021soundstream, defossez2022high} & 1.379 & 3.000 & 0.714 \\
    SQ-RVQ & \textbf{1.443} & \textbf{3.046} & \textbf{0.735} \\
    group SQ-RVQ &  1.401 & 3.034 & 0.730 \\
    SQ $\|$ RVQ & 1.383 & 3.009 & 0.727 \\
    RFSQ \cite{mentzer2023finite} & 1.321 & 2.967 & 0.699 \\
    RLFQ \cite{yu2023language} & 1.330 & 3.038 & 0.703 \\
    \hline
    \end{tabular}  
\end{table}

\begin{table}[t]
    \centering
    \caption{Ablation study on weighted loss function and feature fusion}
    \label{table:oversuppress}
    \begin{tabular}{lccc}
    \hline
     & PESQ $\uparrow$& OVRL $\uparrow$& STOI $\uparrow$\\
    \hline
    RestSE & \textbf{1.443} & 3.046 & \textbf{0.735} \\
    \hline
    $-$ weighted loss & 1.401 & 2.994 & 0.72 \\
    \hline
    $+$ feature fusion & 1.407 & \textbf{3.057} & 0.73 \\     
    \hline
    \end{tabular}  
\end{table}

\begin{figure}[t!]
%\begin{minipage}[b]{1\linewidth}
  \centering
  \includegraphics[width=1\linewidth, height=0.4\linewidth]{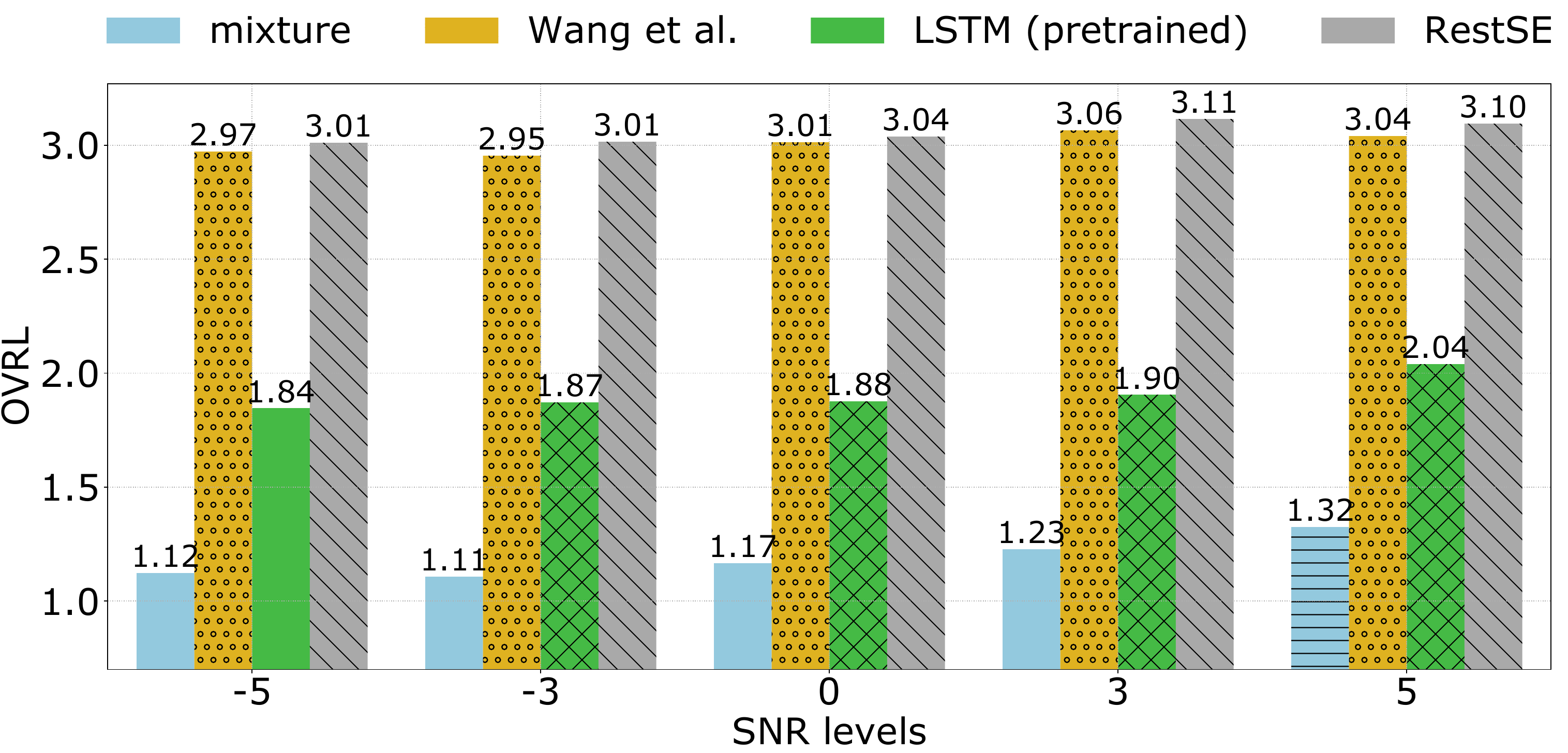} % Adjust width if needed
%\end{minipage}
\caption{Trend bar plot of OVRL scores across various SNR levels.} %(\textbf{Change "codec" to "RestSE", also change the coloar of numbers and text in this figure to black, currently the color is gray.})}
\label{fig:barplot}
\end{figure}

\begin{figure}[t!]
  \centering
  \includegraphics[width=1\linewidth, height=0.43\linewidth]{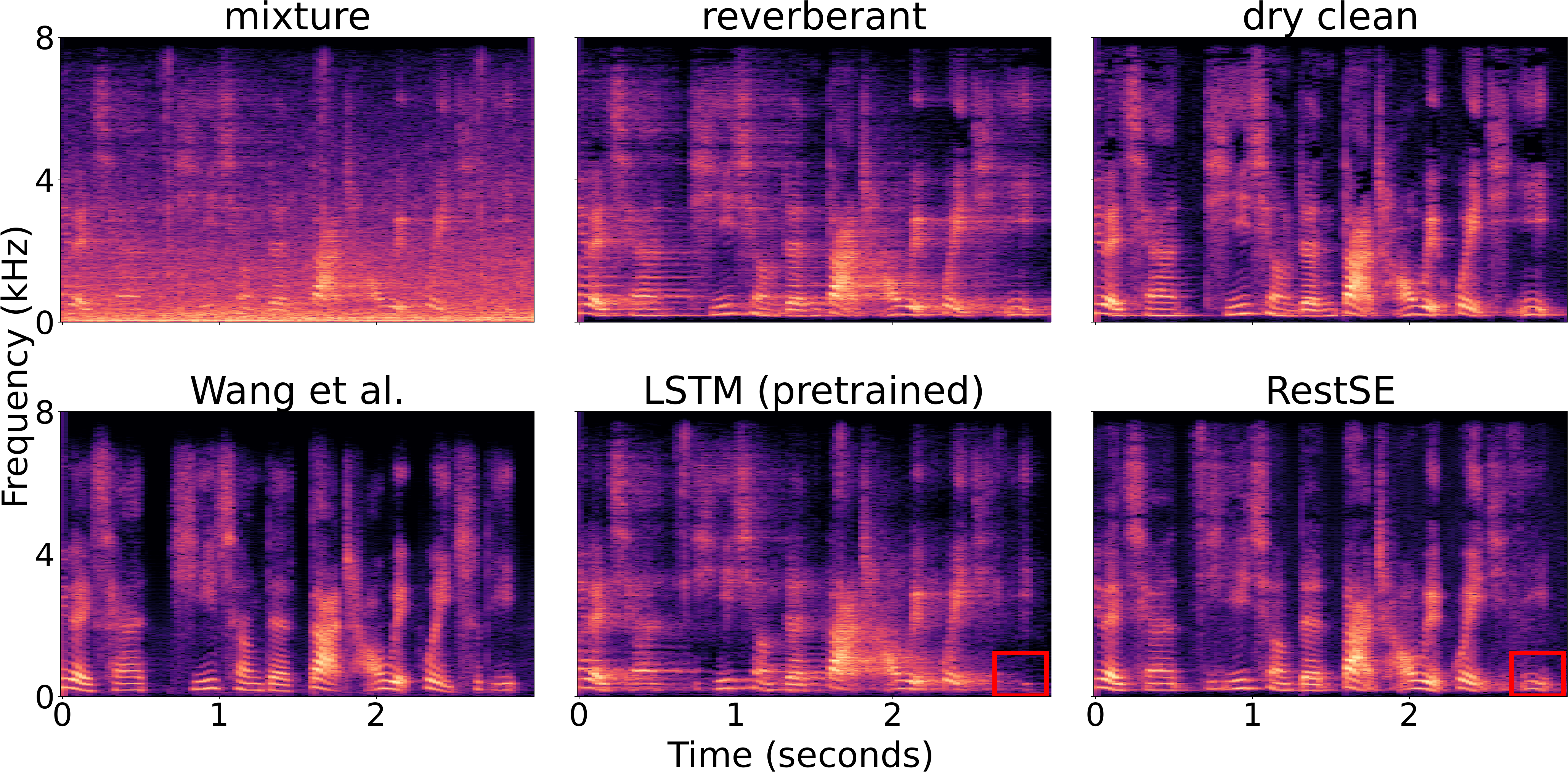}
\caption{Comparison of Spectrograms. The red-boxed areas highlight RestSE's ability to restore regions over-suppressed by the LSTM.} 
\label{fig:spec}
\end{figure}

% \subsection{Evaluation of weighted loss function and feature fusion layer}
Finally, we present a trend bar plot (Fig. \ref{fig:barplot}) and a spectrogram figure (Fig. \ref{fig:spec}) to illustrate the performance of the proposed method.  %(\textbf{Find a proper sample and show the spectrogram of mix, reveb clean, dry clean, LSTM, Heming, RestSE}) to illustrate the performance of the proposed method. 
% The comparison includes the mixture input, Wang et al.'s method \cite{wang2023unifying}, our proposed method using a pretrained LSTM network, and the codec-based approach. Noted that the codec model incorporates the feature fusion layer, as this configuration achieved the highest OVRL score. Fig. \ref{fig:barplot} illustrates the effectiveness of our progressive learning pipeline, where the pretrained LSTM first removes noise, leading to noticeable improvements in OVRL scores. Subsequently, the codec-based approach, which further dereverberates and restores speech, consistently achieves higher OVRL scores compared to Wang et al.'s method. 
% Notably, our method not only outperforms Wang et al.'s work in terms of speech quality but also does so with a smaller model size.
It is seen that our proposed method effectively solves the joint DN and DR problem and consistently outperforms comparison methods.

\section{Conclusion}
In this paper, we propose RestSE - a progressive learning pipeline to enhance speech in noisy and reverberant environments. Our contributions can be summarized as three-fold: first, we design a progressive pipeline where we employ a lightweight SE module to reduce noise while preserving reverberation and utilize a generative codec module for effective dereverberation; second, we explore different quantization techniques, with SQ-RVQ proving most effective; third, we introduce a weighted loss function and a feature fusion approach that merges the SE output with the original mixture to enhance speech restoration. The experimental results demonstrate improvements across objective metrics, highlighting RestSE's potential for challenging speech enhancement tasks.

%\bibliographystyle{IEEEtranS}
\bibliographystyle{IEEEtran}
\bibliography{refs}

\begin{thebibliography}{10}
\providecommand{\url}[1]{#1}
\csname url@samestyle\endcsname
\providecommand{\newblock}{\relax}
\providecommand{\bibinfo}[2]{#2}
\providecommand{\BIBentrySTDinterwordspacing}{\spaceskip=0pt\relax}
\providecommand{\BIBentryALTinterwordstretchfactor}{4}
\providecommand{\BIBentryALTinterwordspacing}{\spaceskip=\fontdimen2\font plus
\BIBentryALTinterwordstretchfactor\fontdimen3\font minus
  \fontdimen4\font\relax}
\providecommand{\BIBforeignlanguage}[2]{{%
\expandafter\ifx\csname l@#1\endcsname\relax
\typeout{** WARNING: IEEEtran.bst: No hyphenation pattern has been}%
\typeout{** loaded for the language `#1'. Using the pattern for}%
\typeout{** the default language instead.}%
\else
\language=\csname l@#1\endcsname
\fi
#2}}
\providecommand{\BIBdecl}{\relax}
\BIBdecl

\bibitem{wang2005ideal}
D.~Wang, ``On ideal binary mask as the computational goal of auditory scene
  analysis,'' in \emph{Speech separation by humans and machines}.\hskip 1em
  plus 0.5em minus 0.4em\relax Springer, 2005, pp. 181--197.

\bibitem{narayanan2013ideal}
A.~Narayanan and D.~Wang, ``Ideal ratio mask estimation using deep neural
  networks for robust speech recognition,'' in \emph{2013 IEEE international
  conference on acoustics, speech and signal processing}.\hskip 1em plus 0.5em
  minus 0.4em\relax IEEE, 2013, pp. 7092--7096.

\bibitem{williamson2015complex}
D.~S. Williamson, Y.~Wang, and D.~Wang, ``Complex ratio masking for monaural
  speech separation,'' \emph{IEEE/ACM transactions on audio, speech, and
  language processing}, vol.~24, no.~3, pp. 483--492, 2015.

\bibitem{lu2013speech}
X.~Lu, Y.~Tsao, S.~Matsuda, and C.~Hori, ``Speech enhancement based on deep
  denoising autoencoder.'' in \emph{Interspeech}, vol. 2013, 2013, pp.
  436--440.

\bibitem{xu2013experimental}
Y.~Xu, J.~Du, L.-R. Dai, and C.-H. Lee, ``An experimental study on speech
  enhancement based on deep neural networks,'' \emph{IEEE Signal processing
  letters}, vol.~21, no.~1, pp. 65--68, 2013.

\bibitem{fu2017raw}
S.-W. Fu, Y.~Tsao, X.~Lu, and H.~Kawai, ``Raw waveform-based speech enhancement
  by fully convolutional networks,'' in \emph{2017 Asia-Pacific Signal and
  Information Processing Association Annual Summit and Conference (APSIPA
  ASC)}.\hskip 1em plus 0.5em minus 0.4em\relax IEEE, 2017, pp. 006--012.

\bibitem{pandey2019tcnn}
A.~Pandey and D.~Wang, ``Tcnn: Temporal convolutional neural network for
  real-time speech enhancement in the time domain,'' in \emph{ICASSP 2019-2019
  IEEE International Conference on Acoustics, Speech and Signal Processing
  (ICASSP)}.\hskip 1em plus 0.5em minus 0.4em\relax IEEE, 2019, pp. 6875--6879.

\bibitem{defossez2020real}
A.~Defossez, G.~Synnaeve, and Y.~Adi, ``Real time speech enhancement in the
  waveform domain,'' \emph{arXiv preprint arXiv:2006.12847}, 2020.

\bibitem{wang2019bridging}
P.~Wang, K.~Tan \emph{et~al.}, ``Bridging the gap between monaural speech
  enhancement and recognition with distortion-independent acoustic modeling,''
  \emph{IEEE/ACM Transactions on Audio, Speech, and Language Processing},
  vol.~28, pp. 39--48, 2019.

\bibitem{pascual2017segan}
S.~Pascual, A.~Bonafonte, and J.~Serra, ``Segan: Speech enhancement generative
  adversarial network,'' \emph{arXiv preprint arXiv:1703.09452}, 2017.

\bibitem{su2021hifi}
J.~Su, Z.~Jin, and A.~Finkelstein, ``Hifi-gan-2: Studio-quality speech
  enhancement via generative adversarial networks conditioned on acoustic
  features,'' in \emph{2021 IEEE Workshop on Applications of Signal Processing
  to Audio and Acoustics (WASPAA)}.\hskip 1em plus 0.5em minus 0.4em\relax
  IEEE, 2021, pp. 166--170.

\bibitem{fang2021variational}
H.~Fang, G.~Carbajal, S.~Wermter, and T.~Gerkmann, ``Variational autoencoder
  for speech enhancement with a noise-aware encoder,'' in \emph{ICASSP
  2021-2021 IEEE international conference on acoustics, speech and signal
  processing (ICASSP)}.\hskip 1em plus 0.5em minus 0.4em\relax IEEE, 2021, pp.
  676--680.

\bibitem{bie2022unsupervised}
X.~Bie, S.~Leglaive, X.~Alameda-Pineda, and L.~Girin, ``Unsupervised speech
  enhancement using dynamical variational autoencoders,'' \emph{IEEE/ACM
  Transactions on Audio, Speech, and Language Processing}, vol.~30, pp.
  2993--3007, 2022.

\bibitem{nugraha2020flow}
A.~A. Nugraha, K.~Sekiguchi, and K.~Yoshii, ``A flow-based deep latent variable
  model for speech spectrogram modeling and enhancement,'' \emph{IEEE/ACM
  Transactions on Audio, Speech, and Language Processing}, vol.~28, pp.
  1104--1117, 2020.

\bibitem{lu2022conditional}
Y.-J. Lu, Z.-Q. Wang, S.~Watanabe, A.~Richard, C.~Yu, and Y.~Tsao,
  ``Conditional diffusion probabilistic model for speech enhancement,'' in
  \emph{ICASSP 2022-2022 IEEE International Conference on Acoustics, Speech and
  Signal Processing (ICASSP)}.\hskip 1em plus 0.5em minus 0.4em\relax IEEE,
  2022, pp. 7402--7406.

\bibitem{lemercier2023storm}
J.-M. Lemercier, J.~Richter, S.~Welker, and T.~Gerkmann, ``Storm: A
  diffusion-based stochastic regeneration model for speech enhancement and
  dereverberation,'' \emph{IEEE/ACM Transactions on Audio, Speech, and Language
  Processing}, 2023.

\bibitem{wang2024selm}
Z.~Wang, X.~Zhu, Z.~Zhang, Y.~Lv, N.~Jiang, G.~Zhao, and L.~Xie, ``Selm: Speech
  enhancement using discrete tokens and language models,'' in \emph{ICASSP
  2024-2024 IEEE International Conference on Acoustics, Speech and Signal
  Processing (ICASSP)}.\hskip 1em plus 0.5em minus 0.4em\relax IEEE, 2024, pp.
  11\,561--11\,565.

\bibitem{richter2023speech}
J.~Richter, S.~Welker, J.-M. Lemercier, B.~Lay, and T.~Gerkmann, ``Speech
  enhancement and dereverberation with diffusion-based generative models,''
  \emph{IEEE/ACM Transactions on Audio, Speech, and Language Processing},
  vol.~31, pp. 2351--2364, 2023.

\bibitem{zhao2017two}
Y.~Zhao, Z.-Q. Wang, and D.~Wang, ``A two-stage algorithm for noisy and
  reverberant speech enhancement,'' in \emph{2017 IEEE International Conference
  on Acoustics, Speech and Signal Processing (ICASSP)}.\hskip 1em plus 0.5em
  minus 0.4em\relax IEEE, 2017, pp. 5580--5584.

\bibitem{zhao2018two}
------, ``Two-stage deep learning for noisy-reverberant speech enhancement,''
  \emph{IEEE/ACM transactions on audio, speech, and language processing},
  vol.~27, no.~1, pp. 53--62, 2018.

\bibitem{wang2023unifying}
H.~Wang, M.~Yu, H.~Zhang, C.~Zhang, Z.~Xu, M.~Yang, Y.~Zhang, and D.~Yu,
  ``Unifying robustness and fidelity: A comprehensive study of pretrained
  generative methods for speech enhancement in adverse conditions,''
  \emph{arXiv preprint arXiv:2309.09028}, 2023.

\bibitem{le2019sdr}
J.~Le~Roux, S.~Wisdom, H.~Erdogan, and J.~R. Hershey, ``Sdr--half-baked or well
  done?'' in \emph{ICASSP 2019-2019 IEEE International Conference on Acoustics,
  Speech and Signal Processing (ICASSP)}.\hskip 1em plus 0.5em minus
  0.4em\relax IEEE, 2019, pp. 626--630.

\bibitem{zeghidour2021soundstream}
N.~Zeghidour, A.~Luebs, A.~Omran, J.~Skoglund, and M.~Tagliasacchi,
  ``Soundstream: An end-to-end neural audio codec,'' \emph{IEEE/ACM
  Transactions on Audio, Speech, and Language Processing}, vol.~30, pp.
  495--507, 2021.

\bibitem{defossez2022high}
A.~D{\'e}fossez, J.~Copet, G.~Synnaeve, and Y.~Adi, ``High fidelity neural
  audio compression,'' \emph{arXiv preprint arXiv:2210.13438}, 2022.

\bibitem{balle2016end}
J.~Ball{\'e}, V.~Laparra, and E.~P. Simoncelli, ``End-to-end optimized image
  compression,'' \emph{arXiv preprint arXiv:1611.01704}, 2016.

\bibitem{yang2024simplespeech}
D.~Yang, D.~Wang, H.~Guo, X.~Chen, X.~Wu, and H.~Meng, ``Simplespeech: Towards
  simple and efficient text-to-speech with scalar latent transformer diffusion
  models,'' \emph{arXiv preprint arXiv:2406.02328}, 2024.

\bibitem{yang2023hifi}
D.~Yang, S.~Liu, R.~Huang, J.~Tian, C.~Weng, and Y.~Zou, ``Hifi-codec:
  Group-residual vector quantization for high fidelity audio codec,''
  \emph{arXiv preprint arXiv:2305.02765}, 2023.

\bibitem{yang2021multi}
G.~Yang, S.~Yang, K.~Liu, P.~Fang, W.~Chen, and L.~Xie, ``Multi-band melgan:
  Faster waveform generation for high-quality text-to-speech,'' in \emph{2021
  IEEE Spoken Language Technology Workshop (SLT)}.\hskip 1em plus 0.5em minus
  0.4em\relax IEEE, 2021, pp. 492--498.

\bibitem{kumar2019melgan}
K.~Kumar, R.~Kumar, T.~De~Boissiere, L.~Gestin, W.~Z. Teoh, J.~Sotelo,
  A.~De~Brebisson, Y.~Bengio, and A.~C. Courville, ``Melgan: Generative
  adversarial networks for conditional waveform synthesis,'' \emph{Advances in
  neural information processing systems}, vol.~32, 2019.

\bibitem{fan2020gated}
C.~Fan, J.~Yi, J.~Tao, Z.~Tian, B.~Liu, and Z.~Wen, ``Gated recurrent fusion
  with joint training framework for robust end-to-end speech recognition,''
  \emph{IEEE/ACM Transactions on Audio, Speech, and Language Processing},
  vol.~29, pp. 198--209, 2020.

\bibitem{zhu2023joint}
Q.-S. Zhu, J.~Zhang, Z.-Q. Zhang, and L.-R. Dai, ``A joint speech enhancement
  and self-supervised representation learning framework for noise-robust speech
  recognition,'' \emph{IEEE/ACM Transactions on Audio, Speech, and Language
  Processing}, vol.~31, pp. 1927--1939, 2023.

\bibitem{du2018aishell}
J.~Du, X.~Na, X.~Liu, and H.~Bu, ``Aishell-2: Transforming mandarin asr
  research into industrial scale,'' \emph{arXiv preprint arXiv:1808.10583},
  2018.

\bibitem{panayotov2015librispeech}
V.~Panayotov, G.~Chen, D.~Povey, and S.~Khudanpur, ``Librispeech: an asr corpus
  based on public domain audio books,'' in \emph{2015 IEEE international
  conference on acoustics, speech and signal processing (ICASSP)}.\hskip 1em
  plus 0.5em minus 0.4em\relax IEEE, 2015, pp. 5206--5210.

\bibitem{allen1979image}
J.~B. Allen and D.~A. Berkley, ``Image method for efficiently simulating
  small-room acoustics,'' \emph{The Journal of the Acoustical Society of
  America}, vol.~65, no.~4, pp. 943--950, 1979.

\bibitem{reddy2022dnsmos}
C.~K. Reddy, V.~Gopal, and R.~Cutler, ``Dnsmos p. 835: A non-intrusive
  perceptual objective speech quality metric to evaluate noise suppressors,''
  in \emph{ICASSP 2022-2022 IEEE International Conference on Acoustics, Speech
  and Signal Processing (ICASSP)}.\hskip 1em plus 0.5em minus 0.4em\relax IEEE,
  2022, pp. 886--890.

\bibitem{mentzer2023finite}
F.~Mentzer, D.~Minnen, E.~Agustsson, and M.~Tschannen, ``Finite scalar
  quantization: Vq-vae made simple,'' \emph{arXiv preprint arXiv:2309.15505},
  2023.

\bibitem{yu2023language}
L.~Yu, J.~Lezama, N.~B. Gundavarapu, L.~Versari, K.~Sohn, D.~Minnen, Y.~Cheng,
  A.~Gupta, X.~Gu, A.~G. Hauptmann \emph{et~al.}, ``Language model beats
  diffusion--tokenizer is key to visual generation,'' \emph{arXiv preprint
  arXiv:2310.05737}, 2023.

\end{thebibliography}

\end{document}